\documentclass[fleqn,twoside]{article}
\usepackage{espcrc2}

\usepackage{graphicx}
\usepackage[figuresright]{rotating}

\newcommand{\AmS}{{\protect\the\textfont2
  A\kern-.1667em\lower.5ex\hbox{M}\kern-.125emS}}

\title{Exact finite-size scaling with corrections in the
       two-dimensional Ising model with special boundary conditions}

\author{W. Janke\address{Institut f\"ur Theoretische Physik, 
        Universit\"at Leipzig,
        Augustusplatz 10/11, 04109 Leipzig, Germany}
        and
        R. Kenna\address{School of Mathematics, Trinity College Dublin, 
        Ireland}
       }
       
\begin{document}

\begin{abstract}
 The two-dimensional Ising model with Brascamp-Kunz boundary conditions
has a partition function more amenable to analysis 
than its counterpart on a torus. This fact is exploited 
to {\em{exactly}} determine the full finite-size scaling behaviour
of the Fisher zeroes of the model. Moreover, exact results are also
determined for the scaling of the specific heat at criticality, 
for the specific-heat peak and for the 
pseudocritical points. All corrections to scaling are found
to be analytic and the shift exponent $\lambda$ does not coincide
with the inverse of the correlation length exponent $1/\nu$.
\vspace{1pc}
\end{abstract}

\maketitle

\section{INTRODUCTION}

Finite-size scaling (FSS) is a well established technique for the 
extraction of critical exponents from finite
volume analyses \cite{bigref}. 
Such exponents characterise
critical phenomena at a second-order phase transition. 
The simplest 
model exhibiting such a transition is the Ising model in two 
dimensions, which, despite a long history and 
extensive study, still offers
new results and insights.
Here, we study the model
under the special boundary conditions 
of Brascamp and Kunz \cite{BrKu74}
to extract new information and 
to help resolve
some hitherto puzzling features of FSS.

Let $C_L (\beta)$ be the specific heat at inverse temperature $\beta$
for a system of linear extent $L$.
FSS of the specific heat is
characterized by the location of its peak,
$\beta_L$,  its height $C_L(\beta_L)$ and 
 its value at the infinite-volume critical point 
$C_L(\beta_c)$. The peak position, $\beta_L$, is a 
pseudocritical point which typically approaches $\beta_c$ as $L 
\rightarrow \infty$ as
\begin{equation}
 |\beta_L - \beta_c| \sim L^{-\lambda}
,
\end{equation}
where $\lambda$ is the shift exponent.
In two dimensions, the Ising specific heat scales as
$
 \ln{L}
$.
Of further interest is the FSS of the  complex 
Fisher zeroes of the partition function \cite{Fi64}.
The leading behaviour of the imaginary part of a Fisher zero
is \cite{IPZ}
\begin{equation}
 {\rm{Im}}z_j(L) \sim L^{-1/\nu}
 ,
\label{IPZI}
\end{equation}
where $z$ stands generically for an appropriate function of temperature,
the subscript $j$ labels the zeroes, and 
$\nu$ is the correlation length critical exponent.
The real part of the lowest zero 
may be viewed as another 
effective critical or 
pseudocritical point, scaling as
\begin{equation}
  | {\rm{Re}}z_1(L) - z_c | \sim L^{-\lambda_{\rm{zero}}}
 ,
\label{IPZR}
\end{equation}
where $z=z_c$ at $\beta = \beta_c$.
Usually the shift exponents, $\lambda$ and $\lambda_{\rm{zero}}$, coincide 
with $1/\nu$, 
but this is not a consequence of FSS and 
is not always true.

The following results have been obtained for FSS in the two-dimensional
Ising model.\\
{\bf{Exact Analytical Results:}}
For toroidal lattices the specific-heat FSS has been determined exactly 
to order $L^{-3}$ at the infinite-volume critical point in
\cite{FF,IzHu00,Sa00}.
Only integer powers 
of $L^{-1}$ occur, with no logarithmic modifications (except 
for the leading term), i.e.,
\begin{equation}
 C_L(\beta_c) =
 C_{00}\ln{L} + C_0
 + \sum_{k=1}^\infty{\frac{C_k}{L^k}}
.
\label{Salas2.13}
\end{equation}
For these periodic boundary conditions
the shift exponent for the specific heat
is  $\lambda = 1 = 1/\nu$,
except for special values of the ratio of the lengths of the lattice
edges, in which case  pseudocritical specific-heat 
 scaling was found to be of the form
$L^{-2}\ln{L}$ \cite{FF}.\\
{\bf{Numerical Results:}}
For spherical lattices the shift exponent of the specific heat was found to be
 significantly away from
$1/\nu = 1$, with $\lambda$ ranging from approximately 
$1.75$ to $2$ (with the possibility of logarithmic corrections)
\cite{spherical}.
Therefore the FSS of the specific-heat pseudocritical point 
does not appear to match the correlation length scaling.

In another  study \cite{ADH}, FSS of Fisher zeroes 
for square periodic lattices yielded a value of $\nu$ which appeared
to approach the exact value (unity) as the thermodynamic
limit is approached. Small lattices appeared to yield an effective 
correction-to-scaling
exponent $\omega \approx 1.8$ while
closer to the thermodynamic limit, these corrections 
tended to be analytic with $\omega=1$.
A certain formal limit of conformal field theory suggests
a correction exponent $\omega = 4/3$ \cite{ZJ}. 
However, the validity of this limit has long been unclear \cite{boris}
and the question of the absence of a subleading operator
corresponding to  $\omega=4/3$ in the standard
Ising model in two dimensions
was recently addressed in depth in \cite{new} (see also \cite{new2}).

In the light of these analyses, we present exact
results which help clarify the situation. To this end, we have selected
the Ising model with Brascamp-Kunz boundary conditions \cite{BrKu74}.

\section{FISHER ZEROES}

The Brascamp-Kunz lattice has $M$ sites in the $x$ direction and $2N$ 
sites in the $y$ direction. 
The boundary conditions are
periodic in the $y$ direction
and the $2N$ spins along the left and right borders
are fixed to $\dots +++ \dots$ and 
$\dots +-+-+- \dots$, respectively.
The partition function is \cite{BrKu74}
\begin{equation}
 Z \propto
 \prod_{i=1}^N{
 \prod_{j=1}^M{
 \left[
         1 + z^2 - z (\cos{\theta_i} + \cos{\phi_j})
 \right]
}}
,
\label{ZM2N}
\end{equation}
where $z=\sinh{2 \beta}$, $\theta_i = (2i-1)\pi/2N$ 
and $\phi_j = j \pi/(M+1)$.
One notes that the partition function (\ref{ZM2N}) is given as a double 
product.
Determination of the Fisher zeroes
of (\ref{ZM2N}) is thus straightforward, as is the calculation of
thermodynamic functions. 
For toroidal boundary conditions, on the other hand,
the partition function is a sum
of four such products \cite{Kaufman}. 
There it is non-trivial to determine
the zeroes or the thermodynamic functions. 

The zeroes of 
(\ref{ZM2N}) are on the unit circle in the complex-$z$ plane (so 
the critical point is $z_c = 1$) \cite{BrKu74}. These are 
$
  z_{ij} = \exp{(i \alpha_{ij})}
$,
where
\begin{equation}
 \alpha_{ij}
 =
 \cos^{-1}{\left( \frac{\cos{\theta_i}+\cos{\phi_j}}{2} \right)}
.
\label{zijM2N}
\end{equation}
One may expand (\ref{zijM2N}) in $M$ to determine the FSS of any
zero to any desired order. Indeed, in terms of the shape parameter
$\sigma = 2N/M$ , the first zero is given by
\begin{equation}
 {\rm{Re}}z_{11} = 
 1
 -
 M^{-2} \frac{\pi^2}{4} \left( 1+ \frac{1}{\sigma^2}  \right)
+ {\cal{O}}\left(M^{-3}\right)
 ,
\label{bb}
\end{equation}
and
\begin{eqnarray}
\lefteqn{{\rm{Im}}z_{11} = 
\frac{ \pi \sqrt{2}}{\sigma (1+\sigma^2)^{5/2}}
\left[
 M^{-1}\frac{(1+\sigma^2)^3}{2}
\right.}
\nonumber\\
& & 
\quad \quad\quad
\left.
 -
 M^{-2}\frac{\sigma^2(1+\sigma^2)^2}{2}
\right]
       +
 {\cal{O}}\left(M^{-3}\right)
 .
\label{cc}
\end{eqnarray}
Higher order terms are straightforward to determine \cite{us}.
 From the leading term in (\ref{cc}) and from (\ref{IPZI}),
the correlation length critical exponent is indeed $\nu =1$. 
Note, however, from (\ref{bb}) that the leading FSS behaviour
of the pseudocritical point in the form of the real part of the lowest 
zero is
\begin{equation}
 z_c - {\rm{Re}}z_{11} = 1 - {\rm{Re}}z_{11}
 \sim M^{-2}
 ,
\label{psze}
\end{equation}
giving a shift exponent $\lambda_{{\rm{zero}}} = 2 \ne 1/\nu$. 
Note further that all corrections are powers of $M^{-1}$ and
thus analytic.

\section{SPECIFIC HEAT}
Since the partition function (\ref{ZM2N}) is multiplicative, the
free energy and hence the specific heat consists of two summations.
These can be performed exactly (we refer the reader to \cite{us}
for details) and one finds the following results.
\\
{\bf{Specific Heat at the Critical Point:}}
At the  critical temperature 
the specific heat is, from (\ref{ZM2N}),
\begin{equation}
 C_{M,2N}^{\rm sing.}(1) =
 \frac{\ln{M}}{\pi}
 \left(
  1 + \frac{1}{M}
 \right)
 +
\sum_{k=0}^\infty{ \frac{c_k}{M^k}}
 .
\label{fsd}
\end{equation}
The coefficients $c_k$ can easily be determined exactly 
and those up to $c_3$ are explicitly given in \cite{us}.
So for the critical specific heat
on a Brascamp-Kunz lattice, 
apart from a trivial $\ln{M}/M$ term (which could be removed
by a redefinition of $M$ \cite{us}),
the FSS
is qualitatively 
the same as (but quantitatively different to)
that of the torus topology in (\ref{Salas2.13}).
\\
{\bf{Specific Heat near the Critical Point:}}
The pseudocritical point of the specific heat, $z_{M,2N}^{\rm{pseudo}}$,
can be determined as the point where the
derivative of $C_{M,2N} (z)$ vanishes. This gives \cite{us}
\begin{eqnarray}
\lefteqn{
 z_{M,2N}^{\rm{pseudo}}
 =
 1
 +
 a_2 \frac{\ln{M}}{M^2}
 + 
 \frac{b_2}{M^2}
}
\nonumber \\
& & 
 + 
 a_3 \frac{\ln{M}}{M^3}
 + 
 \frac{b_3}{M^3}
 + 
 {\cal{O}}\left( \frac{(\ln{M})^2}{M^4} \right)
 ,
\label{pseudocritical}
\end{eqnarray}
higher terms being of the form  $ {\ln{M}}/{M^4} $
and  ${1}/{M^4} $. This implies $\lambda = 2 \ne 1/\nu$
(up to logarithmic corrections).
For the   
specific-heat peak FSS we find \cite{us}
\begin{eqnarray}
\lefteqn{
 C_{M,2N}^{\rm sing.}(z_{M,2N}^{\rm{pseudo}}) = 
 \frac{\ln{M}}{\pi}\left(1+\frac{1}{M} \right)
}
\nonumber \\
& & 
 +
 c^\prime_0
 +
 \frac{c^\prime_1}{M}
 +
 d^\prime_2
        \frac{ (\ln{M})^2 }{M^2}
+
 {\cal{O}} \left(\frac{\ln{M}}{M^2}\right)
 ,
\label{cpeak}
\end{eqnarray}
with $c'_0 = c_0$ and $c'_1 = c_1$.
Higher order terms are of the form 
$1/{M^2}$,
${(\ln{M})^2}/{M^3}$,
${\ln{M}}/{M^3}$
and $1/{M^3}$.
Notice that, up to ${\cal{O}}(1/M)$, (\ref{cpeak}) is quantitatively
the same as the critical specific-heat scaling (\ref{fsd}).
The higher order terms
of (\ref{cpeak})
differ qualitatively from those in
(\ref{fsd}) in that there are  logarithmic modifications 
of the form $(\ln{M})^k/M^l$ (with integer $k$ and $l$).
Again, the values of the coefficients are given in \cite{us}.
           
\section{CONCLUSIONS}
\setcounter{equation}{0}
For the two-dimensional Ising model with Brascamp-Kunz
boundary conditions, we have derived exact expressions for the FSS of
the Fisher zeroes to all orders.
We have also determined the FSS of the critical specific heat,
its pseudocritical point and its peak. The advantage of Brascamp-Kunz
boundary conditions (over periodic ones) is that the partition function is 
a  product and  meliorates determination of higher order corrections.

The following are the main features we have found:
All corrections to scaling are analytic (except for  logarithms).
The shift exponent $\lambda$ does not coincide with $1/\nu$.
The FSS of the specific-heat pseudocritical point and peak
have logarithmic corrections.
Apart from
the leading term, this feature is absent in the critical
specific heat.

\end{document}